# High-contrast Demonstration of an Apodized Vortex Coronagraph

Jorge Llop-Sayson[1], Garreth Ruane[2], Dimitri Mawet[1,2], Nemanja Jovanovic[1], Carl T. Coker[2], Jacques-Robert Delorme[1], Daniel Echeverri[1], Jason Fucik[1], A J Eldorado Riggs[2], and J. Kent Wallace[2]
[1] California Institute of Technology, 1200 E. California Boulevard, Pasadena, CA 91125, USA; jllopsay@caltech.edu
[2] Jet Propulsion Laboratory, California Institute of Technology, 4800 Oak Grove Drive, Pasadena, CA 91109, USA

## Abstract

High-contrast imaging is the primary path to the direct detection and characterization of Earth-like planets around solar-type stars; a cleverly designed internal coronagraph suppresses the light from the star, revealing the elusive circumstellar companions. However, future large-aperture telescopes (>4 m in diameter) will likely have segmented primary mirrors, which cause additional diffraction of unwanted stellar light. Here we present the first high-contrast laboratory demonstration of an apodized vortex coronagraph, in which an apodizer is placed upstream of a vortex focal plane mask to improve its performance with a segmented aperture. The gray-scale apodization is numerically optimized to yield a better sensitivity to faint companions assuming an aperture shape similar to the LUVOIR-B concept. Using wavefront sensing and control over a one-sided dark hole, we achieve a raw contrast of $2 \times 10^{-8}$ in monochromatic light at 775 nm, and a raw contrast of $4 \times 10^{-8}$ in a 10% bandwidth. These results open the path to a new family of coronagraph designs, optimally suited for next-generation segmented space telescopes.

*Unified Astronomy Thesaurus concepts:* Exoplanets (498); Direct imaging (387); Space telescopes (1547)

## 1. Introduction

The formidable task of characterizing atmospheres of nearby worlds with direct imaging techniques is one of the most challenging technological problems in modern astronomy. Future space-based observatories, such as the NASA mission concept of the Large UV/Optical/IR Surveyor (LUVOIR; The LUVOIR Team 2019), include among their primary science goals the detection of molecular species in the atmospheres of exoplanets. However, these exoplanets are many orders of magnitude fainter than their host star; e.g., a rocky planet such as the Earth, orbiting a Sun-like star, requires a starlight suppression system that achieves a contrast of the order of $10^{-10}$ to be imaged. Developments in direct imaging with coronagraph instruments are on the path to providing astronomers with the technologies to tackle these extraordinary science cases.

Many coronagraph designs have been proposed and demonstrated to deal with starlight suppression (Kuchner & Traub 2002; Kasdin et al. 2003; Codona & Angel 2004; Foo et al. 2005; Guyon et al. 2005; Mawet et al. 2005; Soummer 2005; Trauger & Traub 2007). For instance, the vortex coronagraph (Foo et al. 2005; Mawet et al. 2005) is a coronagraph in which an induced azimuthal phase ramp at the focal plane diffracts the starlight toward the outer edge of the beam, where it is blocked at the following pupil plane with a Lyot stop. This concept provides high sensitivity to close-in exoplanets with its enhanced raw contrast and high throughput at small separation angles from the host star. Vortex coronagraphs are widely used on ground-based telescopes, including the W. M. Keck Observatory where it is currently producing high-contrast science with the NIRC2 camera (Serabyn et al. 2017).

However, the number of exoplanets that can be imaged is ultimately limited by the size of the telescope; at a given wavelength the minimum angular separation needed to resolve a planet and its host star is directly proportional to the telescope diameter. Large apertures will be required to undertake the most compelling science cases, e.g., habitable zones of M-type stars with extremely large telescopes from the ground, or a census on rocky planets around solar-type stars with space-based observatories. The next generation of large apertures will be segmented, which will increase the noise from unwanted stellar light due to diffraction from discontinuities in the pupil. The two concepts developed by the LUVOIR concept design team, LUVOIR-A with an on-axis 15 m primary mirror, and LUVOIR-B with an off-axis 8 m primary mirror, both segmented, have their exoplanet science yield driven by their coronagraph performance (Stark et al. 2019), which is greatly affected by the discontinuities in the pupil. Nonetheless, the last decade has seen an array of clever solutions to this problem (Mawet et al. 2011, 2013; Pueyo & Norman 2013; Guyon et al. 2014; Carlotti et al. 2014; Mazoyer et al. 2015; Ruane et al. 2015a, 2015b, 2016; Balasubramanian et al. 2016; Trauger et al. 2016; Zimmerman et al. 2016), and the Exoplanet Exploration Program Office at NASA is currently funding several groups to address this technology challenge through the Segmented Coronagraph Design and Analysis study (Crill & Siegler 2017). For instance, a deformable mirror (DM) assisted vortex coronagraph (DMVC) is baselined for the LUVOIR-B coronagraph, in which the combined work of two DMs suppresses the diffraction from the segment gaps. The same physical outcome in terms of diffraction suppression can be achieved by an apodization of the pupil (Mawet et al. 2013). Ruane et al. (2016) introduced a new family of coronagraph designs, the apodized vortex coronagraph (AVC), in which a vortex coronagraph is modified by apodizing the wavefront at the pupil plane with a gray-scaled pattern optimized to provide an improved sensitivity to close-in exoplanets. The vortex mask and Lyot stop, downstream of the apodizer, suppress the starlight.

Here we show the first laboratory demonstration of an AVC concept using the High Contrast Spectroscopy Testbed for Segmented Telescopes (HCST) in the Exoplanet Technology Laboratory (ET lab) at Caltech. A prototype apodizer was designed and fabricated for a LUVOIR-B type segmented pupil consisting of hexagonal segments with no central obscuration or support struts obscuring the aperture. In Section 2 we present simulations of the expected performance of a LUVOIR-B AVC. In Section 3 we show the laboratory setup at HCST to achieve high levels of contrast and in Section 4 we present the results of the high-contrast demonstrations for monochromatic and broadband light. In Section 5 we discuss the significance of these results for the LUVOIR-B coronagraph, comparing the baselined DMVC to the AVC. Future work and conclusions are discussed in Section 6 and 7, respectively.

## 2. Design and Simulations

We performed simulations comparing the high-contrast performance with and without the optimized apodization. These simulations were performed using the HCST layout, with no other wavefront error other than the pupil discontinuities introduced by the simulated apertures. Simulations of the AVC are performed using the Fast Linear Least-Squares Coronagraph Optimization (Riggs et al. 2018) software package,[3] the same toolbox used to run the HCST.

Our simulations show that the AVC radically improves the starlight suppression within the intended field of view. Figure 1 shows a comparison between the stellar residuals after the coronagraph with (right panel) and without (left panel) the optimized apodization. In theory, a vortex coronagraph provides total rejection of starlight with a flat, evenly illuminated wavefront and a circular aperture. However, the addition of gaps between mirror segments (see Figure 1(a)) causes points of diffracted light to appear throughout the image plane after the coronagraph (see Figure 1(b)) whose brightness depends on the width of the gaps. The gray-scale apodization pattern (see Figure 1(c)) is designed to minimize the diffraction from the star out to an angular separation of ~20 $\lambda/D$ (see Figure 1(d)). The numerical optimization approach is based on Jewell et al. (2017).

In the case of the AVC, the diffraction spikes originating from the presence of the hexagonal segmentation of the pupil are cancelled within the region of interest around the star. The immediate gains in raw contrast are very significant, with an improvement of ~4 orders of magnitude for the AVC in the circular region between 3 and 10 $\lambda/D$ clearly visible in the figure, and a loss in throughput of 8% (for an off-axis source at $6\lambda/D$ from the star).

To emphasize the impact of the apodizer in terms of wavefront control performance, Figure 2 shows the result of two simulations of HCST with a LUVOIR-B type aperture: on the left, without the gray-scale apodization, and on the right, with the AVC. These simulations are for HCST in its two-DM configuration, which allows for a 360° dark hole. The AVC simulation converges to a raw contrast ~2 orders of magnitude better.

## 3. Laboratory Setup

The experiments were performed on the HCST for Segmented Telescopes (Delorme et al. 2018; Llop-Sayson et al. 2019a) in the ET lab at Caltech. HCST is a facility aimed at addressing the technology challenges for high-contrast imaging and spectroscopy of exoplanets with large segmented telescopes. The HCST custom-made optics provide the exquisite wavefront quality required for high-contrast experiments, with <0.016 waves rms (Jovanovic et al. 2018). A custom-made enclosure consisting of sandwiched honeycomb aluminum panels ensures minimum environmental disruption from the exterior, namely air turbulence, acoustic vibrations, and temperature gradient changes. An optical table equipped with active damping isolates the setup from vibrations. The point-spread function (PSF) jitter over a few seconds is 2% $\lambda/D$ rms, and the slower PSF drifts, probably caused by changes in temperature gradients over the testbed, cause drifts from 0.1 to 1 $\lambda/D$ over timescales of a few hours. We work with exposure times of 1 to 5 s and mitigate the effect of the PSF drift by periodically recentering the camera's subwindow.

For the monochromatic light tests we used a laser (Thorlabs S1FC780), while for the broadband tests we used a super-continuum white-light laser source (NKT Photonics SuperK EXTREME) followed by a tunable single-line filter (NKT Photonics SuperK VARIA). The light is fed to HCST through a single mode fiber (Thorlabs SM600); the light from the laser is circularly polarized and reimaged onto a custom-made 5-$\mu$m pinhole.

The layout of the HCST can be seen in Figure 3. The beam is collimated and an iris defines the outer pupil edge to avoid chromatic errors due to vignetting and back-reflection from the apodizer glass substrate prototype. The AO system consists of a DM, or DM (Boston Micromachines Corporation kilo-DM) that controls the wavefront. The DM has a continuous surface membrane with 34 × 34 actuators with an inter-actuator separation of 300 $\mu$m. The apodizer is placed at a pupil plane conjugated with the DM and the entrance iris. After the apodizer, the beam is focused onto the focal plane mask (FPM). HCST uses a vortex coronagraph (Foo et al. 2005; Mawet et al. 2005), which provides an excellent trade-off between the small inner working angle (IWA), throughput, and immunity to low-order aberrations. The vortex coronagraph induces a phase ramp at the focus of the form $e^{\pm il\theta}$, where $l$ is the topological charge of the vortex. Given an arbitrary phase aberration at the pupil plane described as a linear combination of Zernike polynomials, $Z_n^m$, the vortex coronagraph (VC) is insensitive to aberrations such that $|l| > n + |m|$. Here we used a charge $l = 8$ mask, we are thus insensitive of tip and tilt, astigmatism, coma, trefoil, and spherical aberrations. However, a higher charge reduces the throughput at close-in angles, pushing away the IWA. The theoretical IWA for a charge of $l = 8$ VC is ~3.5 $\lambda/D$. A more in-depth analysis of this trade-off can be found on Ruane et al. (2018). After the FPM, the beam is then collimated and clipped at the pupil by the Lyot stop, a circular laser-cut aluminum mask with a 15.4 mm diameter hole that blocks ~93% of the incoming beam diameter. The remaining light is imaged with a ~f/50 beam onto the camera (Oxford Instruments Andor Neo 5.5). In order to do photometric calibration, we used a filter wheel with an neutral density filter when necessary (Thorlabs NE20B, OD = 2.0).

Figure 4 shows the picture of the apodizer prototype used in these experiments, manufactured by Opto-Line. The prototype consists of an AR-coated 6 mm thick BK7 substrate with a microdot pattern on the reflective surface; in this binary mask, the density of microdots on the surface provides the desired

---

[3] https://github.com/ajeldorado/falco-matlab

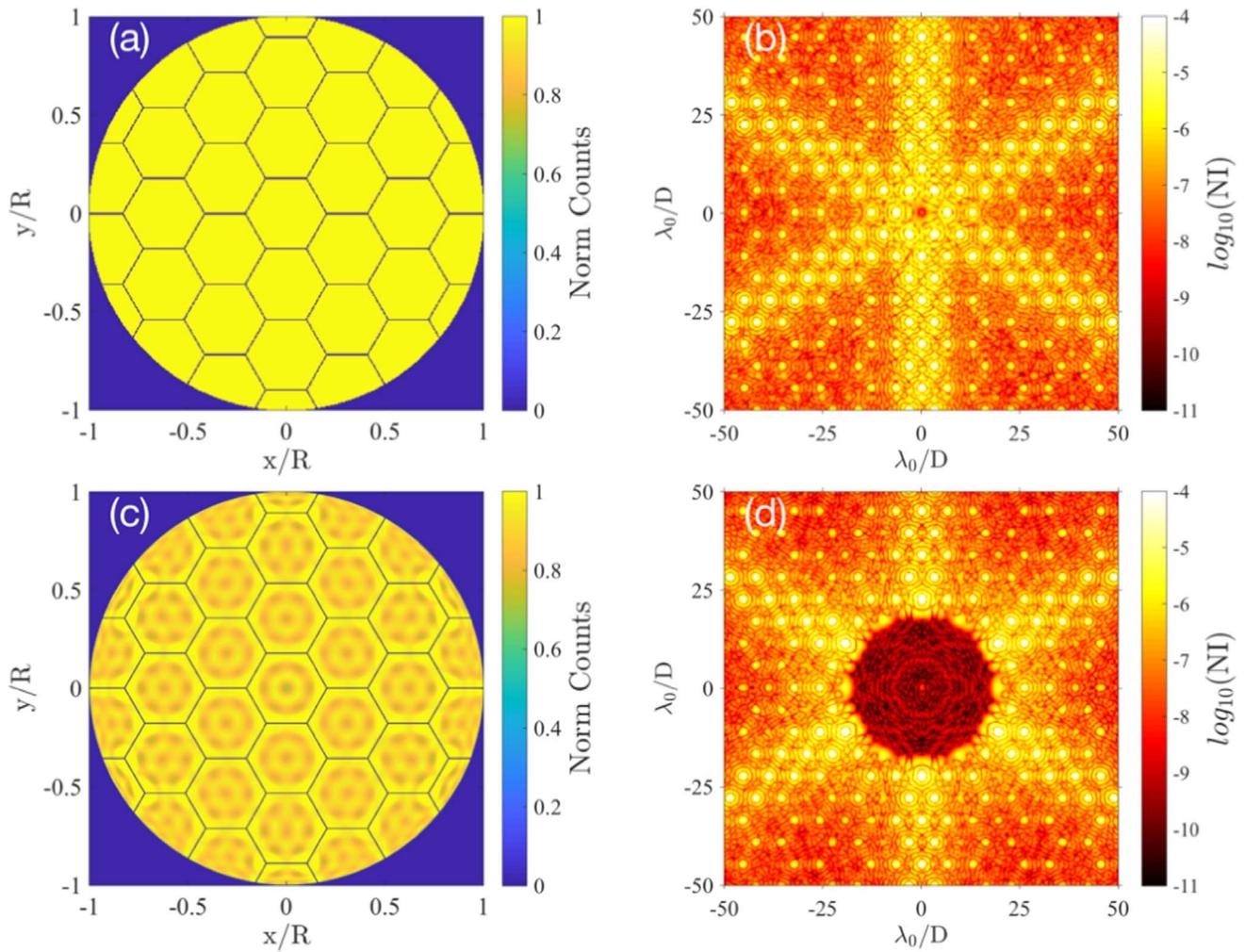

**Figure 1.** Pupil image of a hexagonally segmented LUVOIR-B type telescope aperture (a), with its correspondent simulated stellar coronagraphic PSF (b), and a pupil image of an AVC (c) for the same aperture, with its correspondent coronagraphic PSF (d). The six diffraction spikes are caused by the hexagonal segmentation pattern. No wavefront control has been performed in either case; the dark zone around the center of the PSF for the apodized case is solely due the apodization.

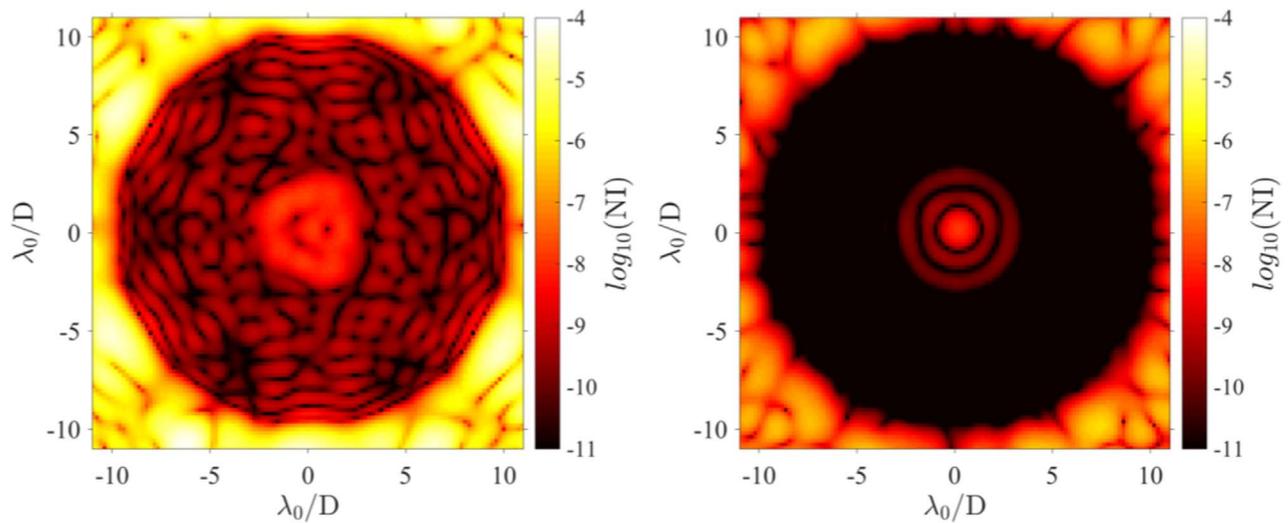

**Figure 2.** Simulated stellar PSFs after wavefront control correction, for a segmented aperture with (right), and without gray-scaled apodization (left).

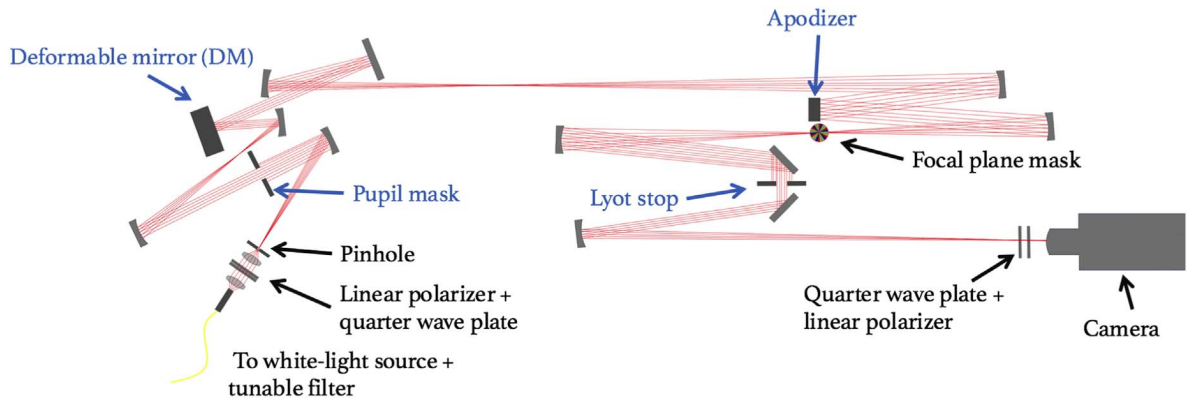

**Figure 3.** Layout of HCST for the apodized vortex coronagraph concept demonstration. Blue font and arrows indicate conjugated pupil planes.

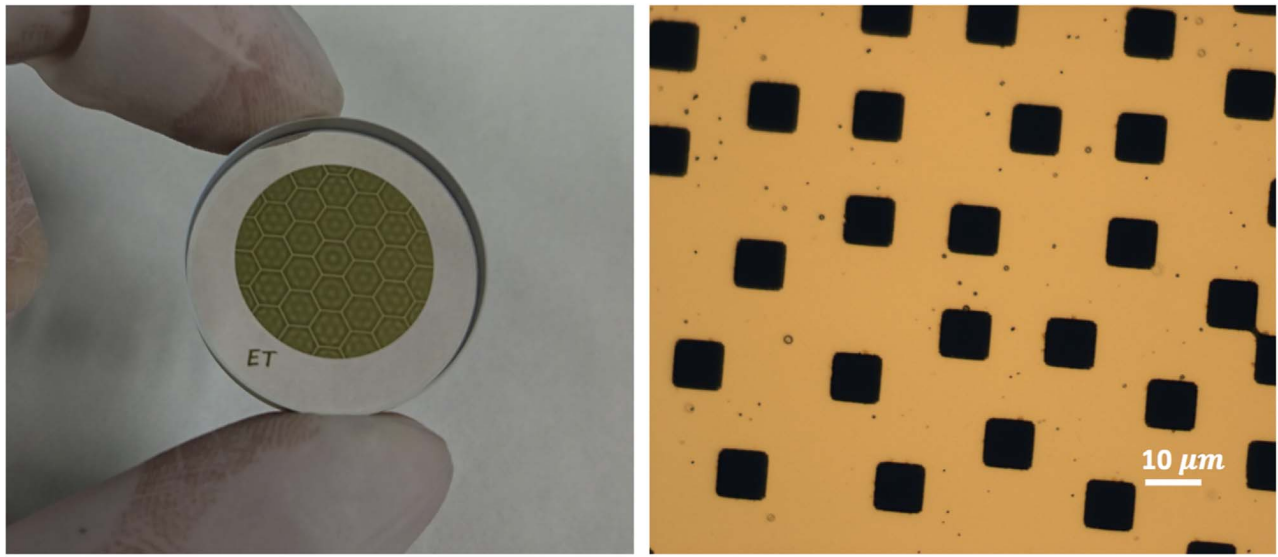

**Figure 4.** Picture of an apodizer prototype (left), and a microscope image of the microdot pattern on the apodizer surface (right). This design is optimized for a LUVOIR-B type aperture, with the gray-scaled apodization achieved with the microdot technique, in which a pattern of $\sim 10 \times 10$ $\mu$m square dots of gold is evaporated onto the substrate surface.

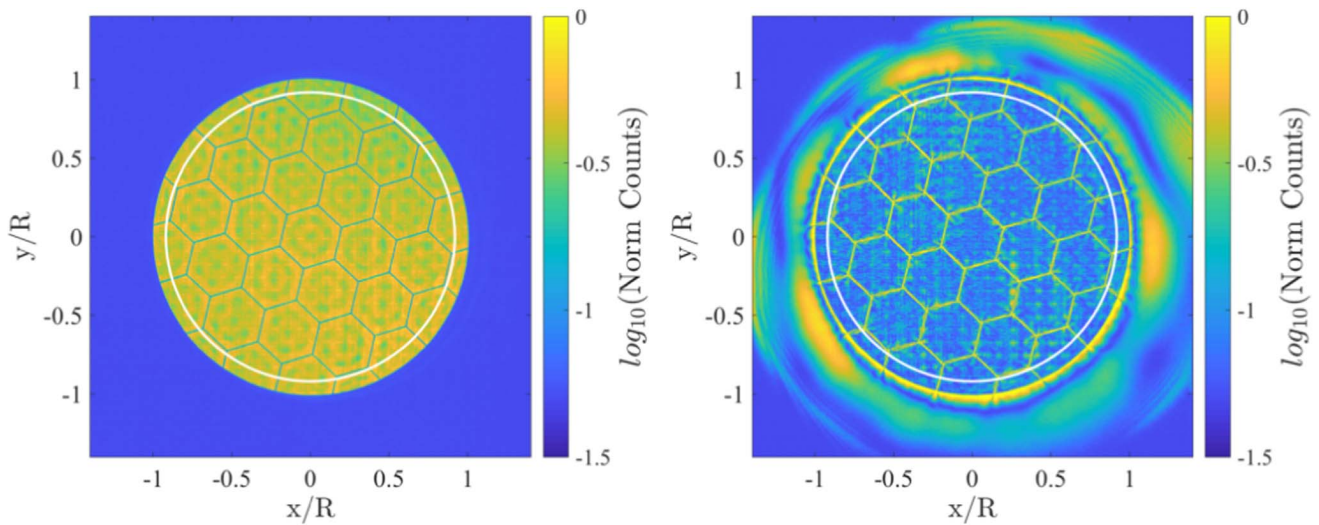

**Figure 5.** Pupil image of the Lyot plane, with focal plane mask out (left) and aligned (right). The white circle indicates the extent of the Lyot stop when aligned with the beam.

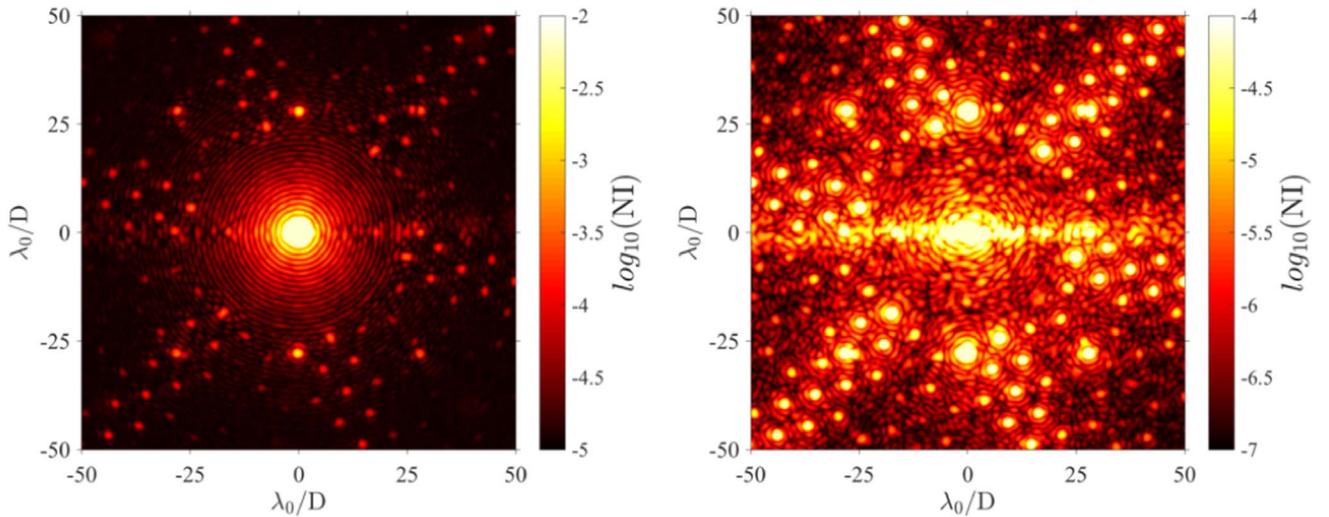

**Figure 6.** AVC PSF (left) and coronagraphic PSF (right). The coronagraphic PSF obtained in the testbed has the same appearance as predicted by simulations in terms of diffraction spikes and apodized area (see Figure 1). Two other major effects can be seen from these images besides the diffraction caused by the hexagonal segmentation: the BMC DM phase error pattern induces a square grid of bright spots at ∼30 $\lambda/D$, and a strong horizontal diffraction stripe can be seen, which is due to phase errors on the OAPs from tooling during fabrication.

gray-scaled apodization. The reflective layer is 400 nm thick with gold evaporated on a thin sublayer of chrome with $10 \times 10\ \mu m$ square voids where the AR coating is exposed.

## 4. Laboratory Results

### 4.1. Results for the AVC in Monochromatic Light

With the laboratory setup described in Section 3 we first demonstrated the AVC concept with monochromatic light. The pupil images with the prototype apodizer aligned in the system are shown in Figure 5, upstream (left) and downstream (right) from the FPM. As expected, with the vortex mask aligned to the beam, the light tends to concentrate in the segment gaps in the pupil downstream of the FPM (Ruane et al. 2018), still this effect is mitigated by the apodization, which aims to send this light out of the beam. The right panel in Figure 5, was taken with the DM turned off (i.e., zero volts applied to the actuators). As such, the azimuthal asymmetry beyond the Lyot stop seen in the image is due to low-order aberrations introduced by the shape of the DM when unpowered. Furthermore, the clipping of the extended beam downstream of the vortex mask is caused by the collimating off-axis parabola (OAP) before the Lyot stop.

In Figure 6 we show the AVC PSF for both an off-axis source and the coronagraphic PSF. The main diffraction effects (other than the Airy ring pattern) that can be identified prior to wavefront control are listed below.

1. The six-fold diffraction spikes are caused by the hexagonal segmentation of the pupil (see Figure 5).
2. The gray-scaled apodization creates a diffraction spike-free area around the simulated star. Without an optimized apodization the diffraction spikes would cover the full field of view and would be difficult to suppress achromatically with the AO system alone.
3. The DM quilting, i.e., the phase pattern on the DM surface, induces a square grid of bright spots at ∼30 $\lambda/D$. This effect is only concerning at levels of raw contrast below $1 \times 10^{-9}$ (Krist et al. 2019).
4. Strong horizontal diffraction features around the simulated star can be seen, which are due to phase errors on the OAP surfaces due to tooling marks at fabrication. Upon inspection with a laser interferometer, all OAPs show vertical stripe-like features with 10 nm rms. The horizontal diffraction is consistent with the surface error measurements.

All major effects before correction with the AO system are thus well understood, namely, the apodizer behaves as predicted creating an area with improved raw contrast (see Figure 1). The starting raw contrast after image sharpening, performed with Zernike tuning with the DM, and with a full-control-area wavefront control run, is below $10^{-6}$ beyond 5 $\lambda/D$.

We performed wavefront sensing and control (WFSC) with the electric field conjugation (EFC; Give'On 2009) algorithm to further suppress residual starlight creating a dark area, or dark hole (DH), around the simulated star. EFC is a model-based algorithm that iteratively finds the DM shape that minimizes the energy in a region of the image plane. It uses a model of the optical system to compute the effect of each DM actuator on the image plane to estimate the electric field on that plane, and to solve for the DM shape that minimizes the energy on the DH. Figure 7 shows the DH image and the resulting DM solution for the correction; the best high-contrast result with laser light is $2 \times 10^{-8}$. In contrast, for previous experiments on the HCST, in which we performed WFSC with EFC with a circular clear aperture, i.e. without the apodizer, we achieved an average raw contrast over a DH of $1 \times 10^{-8}$ for ∼1% of narrowband light (Llop-Sayson et al. 2019a). Although the limitation to HCST's performance with the clear aperture configuration is not fully understood, the most probable cause is a combination of model uncertainty, PSF drift, incoherent light in the system from ghosts, and the limitation from the least significant bit of the DM electronics, which sets the limit of HCST to $7 \times 10^{-9}$ (see Echeverri et al., in preparation). The discrepancy of a factor of two between HCST's best results with and without the AVC could be explained by a combination of a few factors that result from implementing the AVC:

1. *Model uncertainty associated with the apodizer*. For instance, the model mismatch associated with errors from the DM actuator position with respect to the beam may be larger with the apodizer. Indeed, a discrepancy between the DM actuator position relative to the apodizer in the model and the actual relative position in the testbed, considering unaccounted magnification between the two planes, will certainly exacerbate the uncertainty in the model.

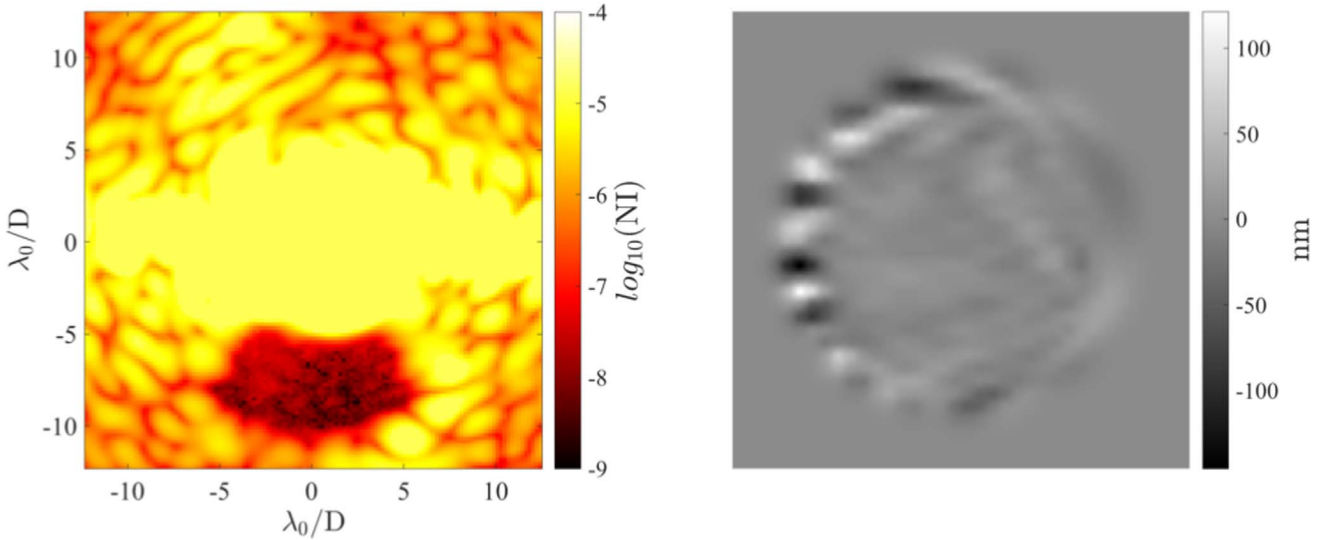

**Figure 7.** Result of an EFC run at HCST with the AVC: coronagraphic PSF with a DH (left), and the corresponding DM solution (right). The best raw contrast achieved is $2 \times 10^{-8}$ with the apodizer prototype used in this experiment; the DH is a 60° aperture arc from 6 $\lambda/D$ to 10 $\lambda/D$. We suspect that the high stroke of the actuators behind the Lyot stop is due to a combination of (1) a positioning error of the Lyot stop in the model with respect to the testbed position, and (2) an effect of the control algorithm dealing with PSF jitter and PSF drift. Here review.

2. *Incoherent light from back-reflection at the back of the apodizer substrate.* Although the substrate of the prototype apodizer is AR-coated, and the beam is circularly clipped at the entrance pupil to match the apodizer circular edge, a small percentage of light is still back-reflected, <1%, and could be an issue at levels of $10^{-8}$ raw contrast.

3. *Lyot plane leakage.* For the clear aperture experiments the Lyot stop would block ∼83% of the radius of the beam, 10% more than for the AVC experiment. This makes leakage at the Lyot plane worse, given that the beam is clipped after the vortex mask (see Figure 5, right image).

4. *Defects in the microdot matrix*(Zhang et al. 2018)*, and/or subtle nonlinear vector diffraction effects due to the subwavelength feature size of microdot edges* (Sivaramakrishnan et al. 2013).

### 4.2. Results for the AVC in Broadband Light

For the broadband demonstration, we chose a 10% bandwidth at 775 nm, and used the NKT VARIA tunable filter to sequentially select equidistant ∼3 nm intermediate bands to perform multiwavelength wavefront control with EFC as in Groff et al. (2016). In Figure 8 we show the result of a corrected coronagraphic PSF with the AVC for broadband light; the best result is of $4 \times 10^{-8}$ average raw contrast for a 60° aperture arc going from 6 $\lambda/D$ to 10 $\lambda/D$ DH.

The average raw contrast for the same DH presented here for the clear circular aperture configuration is currently limited at $3 \times 10^{-8}$ for the same bandpass. As discussed in Section 4.1, different factors associated with the AVC, specifically the apodizer, could explain the discrepancy in the contrast floor.

Furthermore, in the case of broadband light, model errors are harder to trace and tackle.

### 5. Discussion: The AVC versus the DMVC

The LUVOIR-B baseline coronagraph is a DMVC (The LUVOIR Team 2019). A DMVC uses two DMs in series to help suppress the starlight diffracted by the mirror segmentation. Indeed, a two-DM configuration, with both a pupil-plane DM, and an out-of-pupil DM, can correct amplitude discontinuities such as segment gaps. The net remapping effect of the DMVC is strictly equivalent to the gray-scaled apodization of the AVC. The DMVC is all reflective and thus lossless. However, beamwalk on the second out-of-pupil DM makes the DMVC generally more sensitive to low-order aberrations.

The improved robustness to tip and tilt errors for the AVC comes somewhat at the expense of throughput due to the reduced transmittance of the gray-scale apodizer. For a LUVOIR-B like aperture the throughput loss is a marginal ∼9% (Ruane et al. 2018) compared to the DMVC. From the extensive yield analysis of Stark et al. (2019), we found that an AVC on board of LUVOIR-B has an exoEarth yield of 96% of that of the DMVC, which corresponds to a loss of approximately 1 exoEarth. The trade between the sensitivity to low-order aberrations and throughput for LUVOIR-B in terms of the exoEarth yield will be the matter of future work.

Other factors to consider include the associated risk of the DM technology maturity, the appearance of bright spots on the resulting coronagraphic PSF for the DMVC, or the relative alignment error tolerance between the DMs. Furthermore, the DMVC can only deal with a limited segment gap size; the larger the gap, the more DM stroke is needed, and high contrast at the requirement levels of the LUVOIR mission concept, i.e., $10^{-10}$ average raw contrast, is hardly achievable for segment gaps with a thickness of 0.1% of the telescope diameter (Ruane et al. 2018).

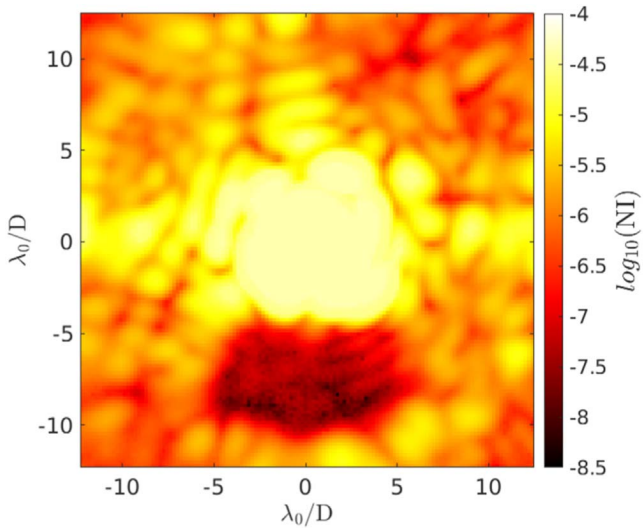

**Figure 8.** Broadband coronagraphic PSF with a DH obtained with EFC. The deepest level of raw contrast achieved at HCST for the AVC with 10% broadband light at 775 nm is $4 \times 10^{-8}$ for a 60° aperture arc from 6 $\lambda/D$ to 10 $\lambda/D$ DH.

## 6. Perspectives

We plan on using a system identification, or system ID (Sun et al. 2018), algorithm based on a neural network to address the model uncertainties in the system. Poorly understood effects at high-contrast levels, such as surface quality and edge effects from the apodizer microdots, or the interplay between actuator positioning in the beam and the segment gaps, could be addressed by this approach. A system ID was successfully implemented at HCST (Llop-Sayson et al. 2019a) and has the potential of dealing with the issues associated with performing model-based WFSC with an AVC, namely the uncertainties coming from the apodizer and segment gaps. At a more general level, demonstrating a system ID for the AVC is directly applicable to any instrument dealing with discontinuities in the pupil of any kind. Such is the case of next-generation extremely large telescopes (ELTs), in which effective model-based WFSC is the pathway to reaching the highest possible number of directly imaged exoplanets.

Plans to improve the performance of HCST are currently underway (Llop-Sayson et al. 2019a), which include: (1) a new source architecture with a new mount, more stable to make the system more robust to PSF jitter and drift, (2) a field stop at the image plane to avoid incoherent light from ghosts, and (3) a tip and tilt sensing and control system. With these upgrades we expect to improve the performance and the limiting factors and thus surpass our current contrast floor.

A fiber injection unit is planned for HCST, with which we will perform WFSC through a single mode fiber (SMF) with the purpose of paving the way for high-dispersion coronagraphy (Sparks & Ford 2002; de Kok et al. 2014; Kawahara et al. 2014; Snellen et al. 2015; Mawet et al. 2017; Wang et al. 2017). Indeed, using an SMF to recover the planet signal improves the sensitivity to planet signal in the presence of starlight noise by virtue of the modal selectivity of the fiber. We previously demonstrated WFSC through an SMF for a clear open aperture (Mawet et al. 2017; Llop-Sayson et al. 2019b), we now plan to use the AVC to demonstrate the capabilities of using an SMF with segmented apertures. Moreover, a custom-made multicore fiber has been purchased to test a multi-object wavefront control approach recently introduced by Coker et al. (2019).

In this paper we have presented an AVC design optimized for a segmentation-only type of aperture, however, although segment gaps are a major concern in coronagraph design, more severe discontinuities, particularly from central obscurations and support struts, pose a more challenging difficulty for high-contrast imaging (Jewell et al. 2017; Stark et al. 2019). Future work will involve efforts on design and testing AVC apodizers optimized for central obscurations and support strut discontinuities.

## 7. Conclusion

We have demonstrated the AVC concept in the laboratory to high levels of contrast with both monochromatic and 10% bandwidth light. The predictions from the AVC model and design process have been validated, as the prototype manufactured for the testbed effectively deals with diffraction emerging from the segmentation from the pupil. Furthermore, WFSC has been successfully implemented with the AVC, consistently reaching levels of $10^{-8}$ raw contrast; for a 60° arc-shaped aperture from 6 $\lambda/D$ to 10 $\lambda/D$ DH, we achieve $2 \times 10^{-8}$ raw contrast for monochromatic light at 780 nm, and $4 \times 10^{-8}$ for a 10% bandwidth at the same wavelength. From previous experiments at HCST with a clear circular aperture, we know that the level of incoherent light is below $1 \times 10^{-8}$ (Llop-Sayson et al. 2019a). We thus plan to address this discrepancy, namely by tackling model uncertainties with a system ID approach and attempting to minimize incoherent light in the system. Furthermore, future experiments with a fiber injection unit will aim to yield improved results in terms of contrast and bandwidth, thus leading the way for future high-dispersion coronagraphy instruments on large segmented telescopes. Indeed, the results presented in this paper, and the envisioned improved performance at HCST with the incoming upgrades, are a testimony of the potential of high-contrast technology in next-generation space-based observatories such as the NASA mission concept LUVOIR, and ground-based observatories, such as the Thirty Meter Telescope.

The first author J.L.S. is partially supported by the National Science Foundation AST-ATI Grant 1710210. Part of this work was carried out at the Jet Propulsion Laboratory, California Institute of Technology, under contract with the National Aeronautics and Space Administration (NASA).

*ORCID iDs*

Garreth Ruane ☉ https://orcid.org/0000-0003-4769-1665
Dimitri Mawet ☉ https://orcid.org/0000-0002-8895-4735
Carl T. Coker ☉ https://orcid.org/0000-0002-9954-7887